\documentstyle[prd,aps,eqsecnum,array,epsf]{revtex}

\begin{document}

\title{Fine-Tuning Solution for Hybrid Inflation in Dissipative Chaotic
Dynamics}

\author{Rudnei O. Ramos$\;
$\thanks{E-mail address: rudnei@peterpan.dartmouth.edu.
$^{\ast\ast}\;$Permanent address.}}

\address{
{\it Department of Physics and Astronomy, Dartmouth College,}\\
{\it  Hanover, New Hampshire 03755-3528, USA}\\
{\it and $^{\ast\ast}$Departamento de F\'{\i}sica Te\'orica,}
{\it Instituto de F\'{\i}sica, Universidade do Estado do Rio de Janeiro,}\\
{\it 20550-013 Rio de Janeiro, RJ, Brazil}}

\maketitle

\begin{abstract} 

We study the presence of chaotic behavior in phase space 
in the pre-inflationary stage of hybrid inflation models.
This is closely related to the problem of initial conditions associated
to these inflationary type of models. We then show how an expected 
dissipative dynamics of fields just before the onset of inflation can solve 
or ease considerably the problem of initial conditions, driving naturally
the system towards inflation.
The chaotic behavior of the corresponding dynamical system is studied 
by the computation of the fractal dimension of the boundary, in phase space,
separating inflationary from non-inflationary trajectories.
The fractal dimension for this boundary is determined
as a function of the dissipation coefficients appearing in the effective
equations of motion for the fields.

\vspace{0.34cm} 
\noindent 
PACS number(s): 98.80 Cq 
\vspace{0.7cm}

\centerline{\large \bf In press Physical Review D}
\end{abstract}

\section{Introduction} 

Among the various proposed models for implementing inflation, the
hybrid inflation model \cite{hybrid} is one of the most attractive ones
due to the possibility of its implementation in the context of 
supersymmetry and supergravity models \cite{super}. 
The principle behind these models for inflation are basically 
based on models where the inflaton is coupled to one or more scalar 
fields. The inflationary phase is characterized by an initial phase in the
evolution of the fields where the inflaton field slowly moves towards
zero vacuum expectation value, till it approaches a critical value
which then induces a spontaneous symmetry breaking in one of the other
fields which are coupled to the inflaton, after which the fields quickly
evolve to their vacuum states and inflation ends.

It has been shown in the context of supergravity motivated models
\cite{supergra}
that for inflaton field amplitudes larger than the Planck mass there
will appear large quantum corrections in the 
inflaton's effective potential, destroying the flatness 
of the inflaton's potential as required by the slow-roll conditions
for inflation. The question then turns to whether inflation
can be achieved in these models for initial inflaton field amplitudes smaller
than the Planck scale, where a reliable model could be constructed.
But in this case it has been shown \cite{fine-tune} that small fluctuations
of the other fields coupled to the inflaton are efficiently able to
prevent the onset of inflation, or quickly drive
previous inflationary trajectories to non-inflationary trajectories and
therefore
making inflation to end before enough e-folds of inflation have been produced
($N_e \gtrsim 70$ as required) to solve the usual cosmological 
horizon and flatness problems. This is related to the homogeneity requirement 
over Hubble size distances \cite{homo} in order to inflation to begin.
In the context of hybrid inflation this problem is even more severe,
requiring an extreme fine-tuning of the initial conditions in order
to have sufficient homogeneity, with negligible spatial and time
derivatives of the other fields that are coupled to the inflaton, over regions
extending far beyond  a few Hubble lengths. 

A few solutions to this homogeneity, or fine-tuning problem, have been 
proposed. The authors in
Ref. \cite{two-stage} for example propose two-stage inflation models
where a first short inflationary phase would smooth a large enough region
in phase space making then possible a later longer second inflationary 
phase. In Ref. \cite{liddle} it is analyzed the possibility of the
solution of the fine-tuning problem in the context of more complicate
generalizations of the hybrid inflation model as motivated from
higher-dimensional scalar field models or brane cosmology.

Here I will be interested in investigating the possible role
of how effective field interactions may shed further light in this
fine-tuning problem in the pre-inflationary phase.  This is mostly
motivated from several works on the effective dynamics of scalar field
configurations, which have shown
that dissipation is intrinsically present in the
field dynamics \cite{GR,morikawa,boya1,boya2,boya3,hu2,hu,BGR}. In
special, the role of dissipation in all of these works has been
emphasized. We can trace the origin of these effects by considering the
dynamics of a given system, which interacts with a sufficiently large
environment (in the sense of degrees of freedom). One well known
example of this, in the context of statistical mechanics,
is the model of an oscillator, taken as
being the system, in interaction with a large number of harmonic
oscillators which are taken as representing the environment. 
By functionally integrating 
over the
environment degrees of freedom we can then show that the dynamics of the
system oscillator will be dissipative, with an equation of motion of the
form of a Langevin-like one \cite{caldeira,hu}.

The dynamic of interacting fields has being a subject of intense
study, in both Minkowski space-time \cite{GR,boya1,BGR}, as also in
FRW space-times \cite{boya2,boya3}, with special interest in
studying the dynamics of the inflaton field both during and immediately
after the inflationary phase. In particular, in Refs. \cite{HR,fang}
it was shown the importance of considering dissipative effects
(due to inflaton's decay) during inflation. The authors in \cite{HR}
have shown how dissipation can influence the usual scenario of
inflation, as also the effect of dissipation on the field trajectories
in phase space.  Strong dissipative regimes for field evolution have
in special motivated the implementation of new inflationary models
called ``warm inflation'' \cite{berera}. The viability of the
construction of microscopic motivated models of strong dissipative
inflaton models has been shown possible in the context of superstring
inspired models in Ref. \cite{BGR2} and  argued also possible for more 
general models in \cite{BR}.

On very general grounds, we are then lead to inquire about the possible
role of field dissipation also at the early stages of inflation and in special
to its role in smoothing large regions in phase space and thus easing
considerable the homogeneity requirements for the onset of inflation
and, consequently, providing a natural solution for the problem of fine-tuning 
associated to hybrid inflation. 
It is easy to understand the effect of dissipation on the initial conditions
problem. {}For the regime of initial inflaton amplitude below the Planck 
scale the energy density is small at the beginning of inflation, resulting
in a small Hubble parameter $H$, which determines how fast the inflaton
energy is converted to expansion. {}For small values of $H$ the coupling
of the fields to the background metric of gravity is small and the 
corresponding friction like term in the equations of motion, of the form
$3 H \dot{\phi}$, is small. This results in a very long evolution for the
fields, where those fields coupled to the inflaton oscillate around 
zero several
times, generating strong sensitivity to very small variations of the 
initial conditions and resulting in a complete indetermination of  
the final outcome
of the fields' trajectories, which can be evolution toward an inflationary
regime or to the minima of the potential. The behavior of the
fields' trajectories in the early times before a required long 
inflationary regime is then extremally chaotic.
In the language of dynamical systems we have two types of attractors
in the system, represented by the inflationary regime and the minima
of the potential. The presence of dissipation works in damping
the fluctuations of the fields in this initial critical time,
suppressing the chaotic motions,
which we then expect, from the results of Ref. \cite{HR},
will bias the inflaton field trajectory toward the inflationary one.   
In fact, recently a study done by the authors in Ref. \cite{ber} 
have indicated that
additional damping terms in the inflaton equation of motion could 
alleviate many of the problems related to the homogeneity requirements
before inflation.  
Another study done by the authors in  Ref. \cite{particle}, 
using a supersymmetric
hybrid inflation model, have also
indicated that particle production may be a way to relax the extreme 
homogeneity requirements for hybrid inflation.

Here we will then be mostly concerned in determining 
for what magnitude
of dissipation will the outcome of the evolution of the fields 
tend mostly toward the inflationary regime, or in the dynamical
system approach, determine how fast dissipation will change the
chaotic regime, characterized by unstable inflationary trajectories, 
to a non-chaotic one, with stable inflationary trajectories.  
We study chaos in our dynamical system of equations by means of the
measure of the fractal dimension (or dimension information) [for a
review and definitions, see {\it e.g.}, Ref. \cite{ott}], which gives a
topological measure of chaos for different space-time settings and it is
a quantity invariant under coordinate transformations, providing then an
unambiguous signal for chaos in cosmology and general
relativity problems in general \cite{levin,levin2}. 
The method we apply in this
work for quantifying chaos is then particularly useful in 
this cosmological pre-inflationary scenario context we are studying,
in which case other methods may be ambiguous,
like, for example, the determination of Lyapunov exponents, which
does not give a coordinate invariant measure for chaos, as discussed in
\cite{levin,levin2}. Also, other methods for studying chaotic systems, like for
example by Poincar\'e sections, are not suitable in the case we are
interested here, in which case chaos is mostly a transitory phenomenon (it ends
by the time the fields reach the potential minima, or when the inflaton
enters the inflationary region).
By computing the fractal dimension characterizing this
chaotic behavior, which is related to the uncertainty in the system parameter
values to predict the final outcome from a given initial condition (we may
say that the boundaries of initial conditions that lead
to inflation or evolution towards the potential minima are mixed) we are
then able to infer the naturalness of inflation for different
settings of the initial conditions. To our
knowledge, this is the first time that this kind of study is performed
in the context of a dissipative dynamical system in cosmology and
applied to the initial condition problem of inflation in particular.

The paper is organized as follows: in Sec. II I give the basic
equations defining the dynamical system and I discuss some of their
properties. In Sec. III
I present the numerical analysis of the dynamical system
and the
computation of the fractal dimension as a function of the magnitude
of dissipation.
{}From this analysis one will be able to conclude about the general effect
of dissipation in the evolution of the fields and how it works in favor
of the inflationary regime.
{}Finally in Sec. IV I give the concluding remarks.

\section{The Model and Its Properties} 

The model I will study here consists of the simplest hybrid 
inflation model, with a scalar 
field $\phi$ (the inflaton) coupled
to another scalar field $\sigma$, which triggers the end of inflation. 
The potential is \cite{hybrid}

\begin{equation}
V(\phi,\sigma) = \frac{\lambda}{4} (\sigma^2-M^2)^2 +
\frac{m^2}{2}\phi^2 + \frac{g^2}{2} \phi^2 \sigma^2  \;,
\label{potential}
\end{equation}

\noindent
where the parameters $M^2,\;m^2$ and the couplings $\lambda$ and $g^2$
are positive. In the following I will treat only the case of 
homogeneous fields, $\phi \equiv \phi(t)$ and $\sigma \equiv \sigma(t)$,
which is the usual case in the works dealing with the initial
condition problem in hybrid inflation.
The interpretation of inflation from the potential (\ref{potential})
is the standard one. {}For values of $\phi$ larger than a critical value
$\phi_{\rm cr}$, where $\phi^2_{\rm cr}= \lambda M^2/g^2$, there is no symmetry
breaking in the $\sigma$-field direction and $\sigma=0$ is a local 
minimum of the potential. We are interested in the region
where $M_{\rm pl}^2 \gg \phi^2 \gtrsim \phi^2_{\rm cr}$ where 
inflation takes place (the false vacuum dominated regime).
After the inflaton field drops below $\phi_{\rm cr}$, symmetry breaking
occurs in the $\sigma$ field direction of the potential.
At this point the fields quickly move towards the minima 
$\sigma=\pm M$, $\phi=0$ and inflation ends.

The chaotic properties of the classical (homogeneous) equations of motion,
in Minkowski spacetime, for a model with potential similar to the one given 
by Eq. (\ref{potential})
were studied in Ref. \cite{bazeia}, while the full effective equations of 
motion
in Minkowski spacetime were studied by Ramos and Navarro in Ref. \cite{RF}.
In Ref. \cite{RF} the general form expected for the 
equations of motion for the fields, for potentials of the form of 
Eq. (\ref{potential}),
was derived and a detailed account for the dissipative
terms appearing in those equations was given. The chaotic behavior of the 
equations of motion for the fields was 
quantified by means of the fractal dimension and the authors studied in details
how dissipation of the fields, due to decaying modes, changed 
the chaotic behavior of the dynamics.
I will here then extend the results of Ref. \cite{RF} to the case
of an expanding background, a flat Friedman-Robertson-Walker background
metric. 

As shown in Refs. \cite{BGR,RF,BR} dissipation comes from the coupling 
of the system fields $\phi$ and $\sigma$ (here taken as background field
configurations) to a bath made of a set of other fields, for 
example made of 
$N_\psi$ fermions $\psi_i$ and/or $N_\chi$ scalars $\chi_i$. The general 
form of the interactions can be of the form 

\begin{equation}
\sum_{i=1}^{N_\chi} (g_\phi^2 \phi^2 \chi_i^2 + 
g^2_\sigma \sigma^2
\chi_i^2)\;,
\end{equation}
for scalar fields $\chi_i$ and 

\begin{equation}
\sum_{i=1}^{N_\psi} (f_\phi \bar{\psi}_i \phi \psi_i + f_\sigma\bar{\psi}_i
\sigma\psi_i)\;,
\end{equation}
for fermion fields $\psi_i$.    
The usual form expected for the coupled effective equations of motion for
$\phi$ and $\sigma$ are complicated non-local equations of motion with
typical non-Markovian dissipative kernels. In \cite{RF} we have shown 
that at high temperature and in the large $N_\chi$, $N_\psi$ limit
for the (thermal) bath fields we can find an approximate Markovian
limit for the kernels, from which we can express the equations of motion
for the fields in terms of dissipative local equations. In Ref. \cite{BR}
we have studied in details the general equations for the 
non-Markovian dissipative kernels at zero temperature and shown
that for a certain class of field decaying modes the Markovian approximation
is also a valid assumption.   
The results obtained in Ref. \cite{BR} show that we can evade the assumption
of an initial high temperature thermal bath as used in Ref. 
\cite{RF} to derive the local 
equations of motion for the fields and makes it, therefore, more appropriate
to the application we intend for here, which is the study of the
initial conditions problem for  the onset of inflation in hybrid inflation
models characterized by potentials like Eq. (\ref{potential}).

I will just use the general form of the coupled effective equations of motion
obtained in Ref. \cite{RF} without a complete specification
of the dissipative coefficients, which we will take as free parameters.
This is a valid analysis here since I will mostly be interested 
in how the general dissipation expected for the fields
$\phi$ and $\sigma$ in their coupled effective equations of motion
will change the chaotic properties of the dynamical system,
which, as explained in the introduction, are direct related to the
initial conditions problem. The detailed form of the dissipative
coefficients depends on the microscopic physics of the specific model
under study, like the coupling of the system fields with the bath fields,
the decaying modes available and the expanding metric. In fact the magnitude
of the dissipation terms may be controlled by these couplings, as shown
in Refs. \cite{BGR,BR,RF}.

In analogy to the results obtained in Ref. \cite{RF} we write the equations of
motion for $\phi$ and $\sigma$ in the form\footnote{This should be 
compared with Eqs. (5.1) and (5.2) of Ref. \cite{RF}. In the notation
used here we have that in those equations that 
$\phi_c \to \phi$, $\psi_c \to \sigma$,
$\bar{m}_\phi^2 \to m^2$, $\bar{m}^2_\psi \to \lambda M^2$,
$\bar{\lambda}_\phi \to 0$, $\bar{\lambda}_\psi \to 6 \lambda$ and 
$\bar{g}\to g$.}

\begin{eqnarray}
\ddot{\phi}  + 3 H \dot{\phi} + m^2 \phi   +  g^2 
\phi \sigma^2  + 
\eta_1 \phi^2  \dot{\phi}  +  
\eta_3 \phi \sigma \dot{\sigma}  =  0 
\label{phi}
\end{eqnarray}

\noindent
and
\begin{eqnarray}
\ddot{\sigma}  +  3 H \dot{\sigma} - \lambda M^2 \sigma   +  
\lambda 
\sigma^3   +  g^2 \sigma \phi^2 + 
\eta_2 \sigma^2  \dot{\sigma}  +   
\eta_3 \phi \sigma \dot{\phi}  =  0 \;,
\label{sigma}
\end{eqnarray}

\noindent
where $\eta_1$, $\eta_2$ and $\eta_3$ denote the dissipation coefficients.
Besides the coupled equations of motion (\ref{phi}) and (\ref{sigma}) we
also have the Friedman equation and the evolution equation for the
Hubble parameter, $H = \dot{a}/a$, where $a$ is the scale factor,
are given as usual by

\begin{equation} 
H^2 = \frac{8 \pi G}{3} (\rho_m+ \rho_r) -
\frac{k}{a^2} \;, 
\label{G1} 
\end{equation} 

\begin{equation} 
2 \dot{H} + 3 H^2 + \frac{k}{a^2} = - 8 \pi G (p_m +
p_r) \;, 
\label{G2} 
\end{equation}

\noindent 
where $G= 1/m_{\rm
Pl}^2$, with $m_{\rm Pl}$ the Planck mass. $k=0,+1,-1$ for a flat,
closed or open Universe, respectively. $\rho_{m (r)}$ and
$p_{m (r)}$ are the energy density and pressure for matter
(radiation), respectively. We also have the standard relations:

\begin{equation}
\rho_m = \frac{1}{2} \dot{\phi}^2 
+\frac{1}{2}\dot{\sigma}^2+ V(\phi,\sigma)\;,
\label{rhom}
\end{equation}

\begin{equation}
p_m = \frac{1}{2}
\dot{\phi}^2 +\frac{1}{2}\dot{\sigma}^2
- V(\phi,\sigma)
\label{pm}
\end{equation}

\noindent
and $p_{r} = \frac{1}{3} \rho_{r}$.

The matter and radiation energy densities $\rho_m$ and 
$\rho_r$ evolve in time as: 

\begin{equation} 
\dot{\rho}_m + 3 H \left(\dot{\phi}^2 + \dot{\sigma}^2\right)+ \eta_1
\phi^2 \dot{\phi}^2 + \eta_2 \sigma^2 \dot{\sigma}^2 + 2 \eta_3 \phi \sigma
\dot{\phi} \dot{\sigma}=0 \;
\label{drhom} 
\end{equation} 

\noindent
and (from the energy conservation law)

\begin{equation} 
\dot{\rho}_r + 4 H \rho_r - \eta_1 \phi^2 
\dot{\phi}^2 -\eta_2 \sigma^2 \dot{\sigma}^2 - 2 \eta_3 \phi \sigma
\dot{\phi} \dot{\sigma}=0 \;. 
\label{drad} 
\end{equation}

Assuming a flat universe ($k=0$), from Eq. (\ref{G1}), we can consider 

\begin{equation}
\rho_r = \frac{3}{8 \pi G} H^2 - \rho_m =
\frac{3}{8 \pi G} H^2 - \frac{\dot{\phi}^2}{2} -  \frac{\dot{\sigma}^2}{2}
- V(\phi, \sigma) 
\label{sol rhor}
\end{equation}

\noindent
as the first integral of Eq. (\ref{drad}). Using Eqs. (\ref{G1})-(\ref{pm}), 
we can also express the equation for the acceleration in the following form

\begin{equation}
\frac{\ddot{a}}{a} = \frac{8 \pi G}{3} (\rho_m - \rho_r)-
4 \pi G\left(\dot{\phi}^2 + \dot{\sigma}^2\right)\;.
\label{G3}
\end{equation}

\noindent
The field equations (\ref{phi}) and (\ref{sigma}) with (\ref{G3})
form a dissipative dynamical system that we will study numerically.
Using (\ref{sol rhor}) in (\ref{G3})
and defining the following dimensionless
variables: $y = \sqrt{8 \pi G} \phi$, $x= \sigma/M$, $\tau = M t$ and 
the constants $\alpha^2 = m^2/M^2$, $\beta^2 = 8 \pi G M^2$ and
the rescaled dissipative coefficients $\eta_y = M \eta_1$, $\eta_x=
M \eta_2$ and $\eta_{xy}= M \eta_3$, the system of equations
(\ref{phi}), (\ref{sigma}) and (\ref{G3}) can then be rewritten in terms
of dimensionless variables  in the form of the following system of first order 
differential equations:

\begin{eqnarray}
&&x' = z \;,\nonumber\\
&&y'=w \;,\nonumber \\
&&a'=u\;,\nonumber \\
&&z' = -3 \frac{u}{a} z + \lambda x - \lambda x^3 -
\frac{g^2}{\beta^2} x y^2 - \eta_x x^2 z - \frac{1}{\beta^2} \eta_{xy}
x y w \;,\nonumber \\
&&w' = -3 \frac{u}{a} w - \alpha^2 y - g^2 y x^2 - \frac{1}{\beta^2}
\eta_y y^2 w - \eta_{xy} x y z \;,\nonumber\\
&&u'=-\frac{u^2}{a}-\frac{a}{6} \left(\beta^2 z^2+w^2-\beta^2 \lambda +
2 \lambda \beta^2 x^2 - \lambda \beta^2 x^4 - 2 \alpha^2 y^2 - 2 g^2 x^2 y^2
\right)\;,
\label{system}
\end{eqnarray}

\noindent
where prime indicates derivative with respect to the dimensionless 
time $\tau$, {\it e.g.}, $x'= dx/d\tau$.

{}For convenience we use the values for $\lambda$, $g^2$, $M$ and $m$
as given by Mendes and Liddle in Ref. \cite{liddle}: $\lambda = g^2 =1$,
$M=2 \times 10^{-2} M_{\rm Pl}$ and $m=5 \times 10^{-6} M_{\rm Pl}$,
where $M_{\rm Pl} = 1/\sqrt{8 \pi G}$ is the reduced Planck mass.
{}From these values of $m$ and $M$ we find for the 
constants $\alpha$ and $\beta$ in Eq. (\ref{system}) the values:
$\alpha= 2.5 \times  10^{-4}$ and $\beta= 2 \times 10^{-2}$. 
In terms of these values we also have that the critical value for the
inflaton field, $\phi_{\rm cr}$, in the dimensionless variables, 
is given by $y_{\rm cr} = 2 \times 10^{-2}$.

\section{Chaos and the Fractal Dimension}

We next numerically solve the system of equations in Eq. (\ref{system})
and we search for chaotic regimes and  how they change with increasing
dissipation. Lets initially consider the following
 initial conditions at $t=0$:
$\phi= 4 \times \phi_{\rm cr}$, $\dot{\phi}=0$, $\sigma=
2\times 10^{-2} M$, $\dot{\sigma}=0$ and $H$ is determined initially
by Eq. (\ref{G1}), with zero initial radiation energy density.
These values guarantee that the stable inflationary trajectories will
correspond to at least $N_e \gtrsim 60$ e-folds of inflation.
{}For the dissipation coefficients $\eta_x$, $\eta_y$ and $\eta_{xy}$,
for convenience, 
we will consider them all with the same value, which is consistent
with the recent findings in Ref. \cite{BR}, or the calculations
in Ref. \cite{RF}. Small variations of the magnitude
among these coefficients are not critical here, since the dissipation in
Eq. (\ref{system}) is dominated by the term corresponding to $\eta_1 
\phi^2 \dot{\phi}$. We take the dissipation coefficients with
initial values as given by $\{\eta\} = (\eta_x,\eta_y,\eta_{xy})=
(0,0,0)$ and increase them till the chaotic behavior of the
system changes to non-chaotic. 

{}For $\{\eta\}=0$ a typical result for three trajectories around
the initial values of the fields (in dimensionless units), separated
by a $\pm 10^{-5}$ variation, is shown in {}Fig. 1.

In {}Fig. 1 we can distinguish the inflationary trajectory represented
by the straight line at $x=0$, which represents the evolution of
the fields in phase space along the valley of local minima
of the potential, $\sigma=0$ for $\phi >\phi_{\rm cr}$. This trajectory
is characterized by a very long evolution by which the universe 
expands exponentially. The number of e-folds produced by the end of the
inflationary phase, when the inflaton reaches $\phi \lesssim \phi_{\rm cr}$,
for this particular trajectory is $N_e \sim 350$. 
The other two oscillatory trajectories, 
that evolve towards $x=\pm 1$ represent the non-inflationary
trajectories, where the fields quickly evolve towards
the potential minima $\sigma=\pm M$ and $\phi=0$. 
In {}Fig. 2 we show a blow-out of the initial time evolution
of Fig. 1. It clearly shows highly oscillatory chaotic like
behavior.

The chaotic behavior of the
dynamical system is quantified  by means of the determination of
the fractal dimension of the boundary separating the inflationary 
trajectories from those that are not inflationary, {\it i.e.},
the ones that evolve towards the minima of the potential,
$\phi=0$ and $\sigma =  \pm M$.
The fractal dimension is associated with the possible different exit
modes under small changes of the initial conditions at $t=0$
and it will give a measure
of the degree of chaos of our dynamical system. The exit modes
we refer to above are ones of the symmetry breaking minima in the
$\sigma$-field direction, $\sigma=\pm M$, $\phi=0$ and the inflationary
one, which are attractors of field trajectories in
phase space. The method we employ to determine the fractal dimension is
the box-counting method which is a standard method for determining
the fractal dimension of boundaries \cite{ott}. Its definition and the 
specific numerical
implementation we use here have been described in details in Ref.
\cite{fractal}.

The basic procedure is, given a set of initial conditions ${\bf x}_0$
at $t=t_0$, which leads to a certain outcome for the trajectory in
phase space, by perturbing them by an amount $\delta$ we then
study whether there will be a change of outcome for the trajectory or
not (whether the perturbation will lead to a different attractor or 
not). Given a volume region in phase space around a boundary 
between different attractors and perturbing a large set of initial 
conditions inside that region, the fraction
of uncertain trajectories, $f(\delta)$, which result in a 
different outcome under a small perturbation, can be shown to scale
with the perturbation $\delta$ as \cite{ott}
$f(\delta) \sim \delta^\epsilon$,
where $\epsilon$ is called the uncertainty exponent. The box-counting
dimension of the boundary in phase space separating different
attractors, or fractal dimension $f_d$, is given by~\cite{ott}
$f_d = d -\epsilon$, where $d$ denotes the dimension of the phase 
space. {}For a
fractal boundary $f_d > d -1$, implying that $\epsilon < 1$, whereas for a
non-fractal boundary, $f_d = d - 1$, and $\epsilon = 1$.

{}Following the method of box-counting we then consider a box in phase space 
(for the dimensionless
variables) of size $10^{-5}$,
around the initial
conditions used in {}Figs. 1 and 2, 
inside which a large number of random
points are taken (a total of $100.000$ random points were used in each
run). All initial conditions are then numerically evolved by using an
eighth-order Runge-Kutta integration method and the fractal dimension is
obtained by statistically studying the outcome of each initial condition
for each run of the large set of points. Special care is taken to keep
the statistical error in the results always below $\sim 1 \%$. 
{}From these numerical simulations we obtain for the zero dissipative
regime of {}Fig. 1 the result for the fractal dimension as given by
$f_d \simeq 5.80 \pm 0.05$. This corresponds to the dimension of the
dynamical system, Eq. (\ref{system}), corresponding here to $d=6$,
minus the uncertainty coefficient
$\epsilon$ (which gives a measure of how chaotic is the system 
\cite{fractal}), $f_d= 6 - \epsilon$.
The results obtained by increasing the value of the dissipation
coefficients $\{\eta\}$, while keeping the same initial
condition used above, around which the perturbations are
taken, are shown in {}Fig. 3.

The results in {}Fig. 3 show that the system quickly changes to a non-chaotic
behavior for a dissipation coefficient around 
$\{\eta\}\sim 3.1 \times 10^{-4}$. {}For those values of dissipation and
higher all trajectories are inflationary ones with $N_e \gg 60$.
At that point we then have the breakdown of the fractal structure 
of the boundary between inflationary and non-inflationary trajectories.
The boundary becomes smooth and we can reliable predict that the initial
conditions will evolve towards the inflationary region.

In order to study how efficient is dissipation to drive non-inflationary
trajectories towards the inflationary region, which is characterized 
by $\ddot{a} >0$, we show in {}Fig. 4 the results for the number of e-folds
of inflation as a function of dissipation for an initial trajectory 
that in the absence of dissipation would be a non-inflationary one.
The particular initial conditions we take purposefully 
correspond to fields well below the Planck
scale, 
$\phi=3.5 M$, $\sigma=M$, and non-vanishing field derivatives
$\dot{\phi}= 2.5\times 10^{-2} M^2$ and $\dot{\sigma}=5 \times 10^{-4} M^2$,
which in the absence of dissipation would be quickly driven to one of the 
minima of the potential (and therefore corresponding to a non-inflationary
behavior). 
We see that above some value of dissipation coefficients ${\eta}\sim 1$
the number of e-folds of inflation increases fast for rather very small
increase of dissipation. We then clearly see using the set
of initial conditions above how dissipation acts in turning 
non-inflationary trajetories into inflationary ones, shifting the system
deep into the inflationary region as dissipation is increased.
The initial radiation energy density in this case, 
as in the previous one, is taken initially zero and remains always smaller 
than the vacuum energy density during the whole evolution of the inflaton.
In particular, for the values of dissipation coefficients we have 
studied here, after a few e-folds of inflation the radiation
energy density is a negligible fraction of the vacuum energy density
and right before the end of inflation the radiation energy
density is a tiny fraction of the vacuum energy density, $\rho_r/\rho_v
\sim 10^{-14}$. A detailed treatment of the liberated radiation to the
medium and its eventual translation to a temperature depends of course
on how the radiation thermalization rate $\Gamma_{\rm rad}$, which
is determined by the microscopic physics,
compares to the expansion rate $H$.

It is also useful to compare the regimes of field dissipation
corresponding for instance to the ones studied in {}Fig. 3 and 
{}Fig. 4 and their relation to the warm inflation scenario. In 
warm inflation \cite{BGR,BGR2} we are usually interested in the 
regime of strong
dissipation in the effective equation of motion
for the inflaton, as compared with the expansion term $3 H \dot{\phi}$.
{}From Eq. (\ref{phi}) dissipation is dominated by
the term $\eta_1 \phi^2  \dot{\phi}=\eta(\phi) \dot{\phi}$.
{}From the values of parameters and initial conditions used in {}Fig. 3 
we find that
$3 H/\eta(\phi) \sim 5$ for the largest value of dissipation studied,
while for the parameters and initial conditions used in {}Fig. 4 we 
have that $3 H/\eta(\phi) 
\sim 0.008$ and therefore deep in the regime of warm inflation.
We can easily see then that warm inflation kind of models does not
seem to suffer from the initial conditions problem seen in standard hybrid
inflation models. In particular, for the parameters used in {}Fig. 3,
for values of dissipation ${\eta}\gtrsim 0.0015$ we enter in the warm
inflation regime and the system has well before 
became non-chaotic, with stable inflationary
trajectories.

\section{Conclusions} 

We have examined a dynamical system describing hybrid inflation 
in the presence of dissipation for both the inflaton and the 
field that triggers the end of inflation. Dissipation is taken as
being present throughout the system's evolution.
In the absence of dissipation the dynamical system is shown to
be highly chaotic by means of the measure of the fractal dimension.
The fractal dimension gives a coordinate invariant measure for chaos 
and therefore is an appropriate way to quantify chaos in cosmology. 

{}For a given initial condition that evolves to an inflationary region, 
by varying it by a small perturbation we completely loose track of its final
outcome, which may still evolve towards the inflationary region,
characterized by evolution along the valley of local minima of the potential,
characterizing the inflationary regions, or fast evolution
towards one of the symmetry breaking minima. The time evolution of
the fields are 
shown to be highly oscillatory for a very long time, which 
eventually change the inflationary trajectories to non-inflationary ones.
This is the main characteristics in the evolution of the fields and that
gives rise to the initial condition problem in hybrid inflation
(and in a lesser degree in any inflationary model). The long oscillatory 
behavior of the fields is highly chaotic. By measuring the magnitude
of chaos in the system we can then infer the severity of the 
initial conditions problem. 

However, by changing the magnitude of dissipation we have then shown 
that the highly chaotic evolution of the system, with unstable
inflationary trajectories, can change quickly
to a non-chaotic, stable one, where all trajectories eventually evolve
toward the inflationary region.
Dissipation acts by damping the initial oscillatory behavior of
the fields and then making it possible to provide a solution for 
the initial condition
problem. We have shown that this solution can be obtained with
relative very small dissipation of the initial inflaton amplitude,
with rates of conversion of vacuum energy to radiation as
small as $\sim 10^{-14}$.
Here dissipation works in two ways. By damping the fluctuations 
associated to the fields coupled to the inflaton the
fine-tuning problem associated with hybrid inflation models is avoided
and at the
same time it makes possible for initial inflationary trajectories
to enter faster in the inflationary region (characterized by $\ddot{a}
>0$) and stay longer in there.

The method used here to study the initial condition problem
can also be easily extented to include other variants of the
hybrid inflation model, for example in the context of 
supersymmetric motivated potentials, where more than one
field is coupled to the inflaton. The additional equations
of motions will also generically be dissipative equations
and the same changing of behavior due to dissipation
observed here is also expected to appear in those more
complicated models.

There have been previous studies on chaos in hybrid inflation models
\cite{maeda,bruce,bellido}, but these studies have concentrate on the
reheating period after the inflationary phase. In that context, the
effect of dissipation, that would be inherently present in the field
equations also during this period, has been neglected. It
would then 
be interesting to investigate in those cases also the effect of dissipation 
and its role, for example, in the phenomenon of parametric resonance 
during preheating
in hybrid inflation models \cite{brand,sanderson}.
We are currently investigating this and other applications
and we expect to report on them soon.

\acknowledgments

I would like to thank R. Caldwell and M. Gleiser for many discussions in
the subject of this paper and on the applications of dissipative regimes for
scalar fields in cosmology. The author was supported by Conselho Nacional de
Desenvolvimento Cient\'{\i}fico e Tecnol\'ogico - CNPq (Brazil)
and SR-2 (UERJ).

\newpage

\begin{figure}[b] 
\epsfysize=9cm  
{\centerline{\epsfbox{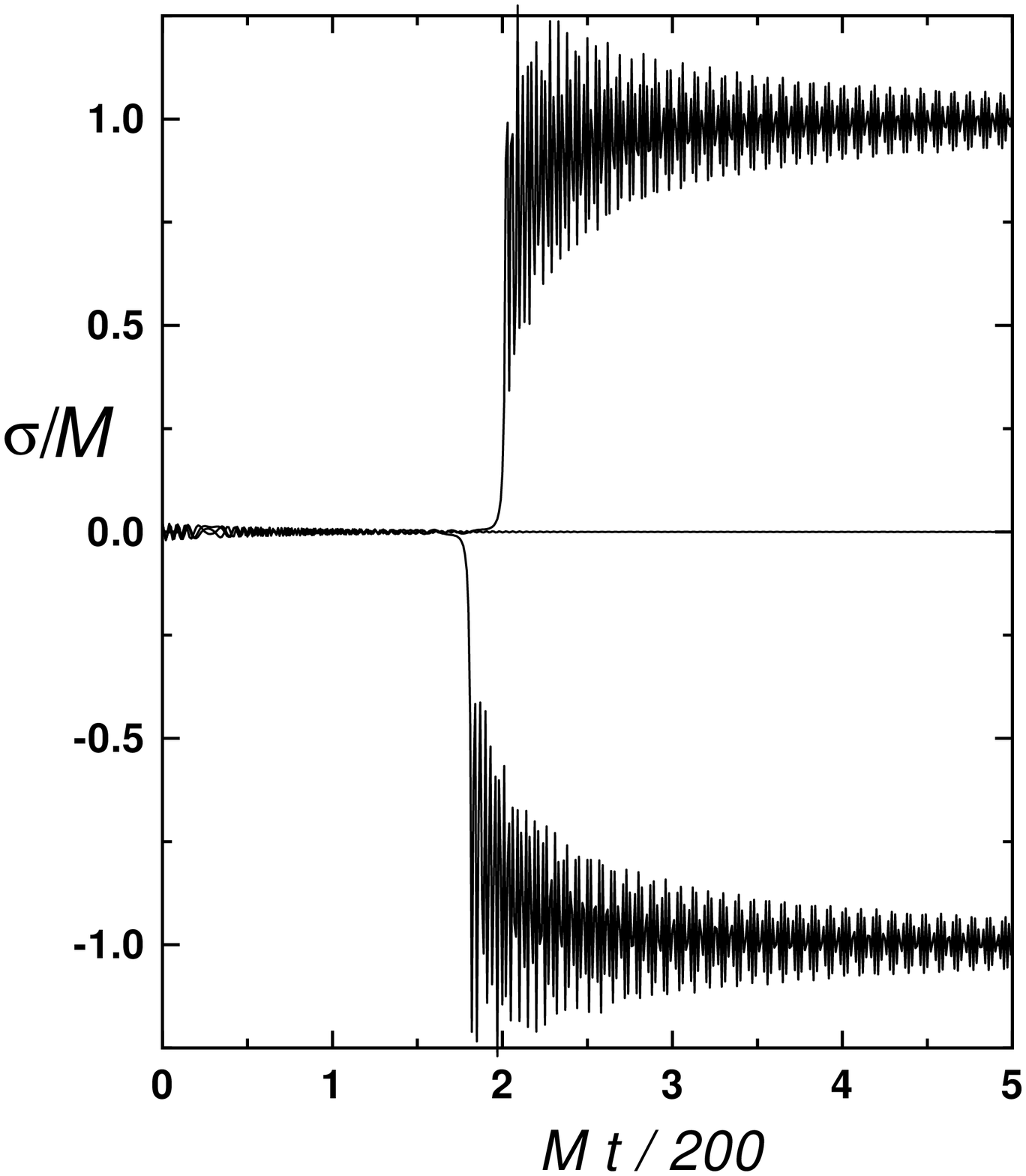}}}
\caption{A typical evolution of three trajectories in the absence of
dissipation with initial conditions
separated by a $\pm 10^{-5}$ variation around the $t=0$ values
$\phi= 4 \times \phi_{\rm cr}$, $\dot{\phi}=0$, $\sigma=
2\times 10^{-2} M$, $\dot{\sigma}=0$ and $H$ determined
by Eq. (\ref{G1}).}
\vspace{0.5cm} 
 
\end{figure} 

\begin{figure}[b] 
\epsfysize=9cm  
{\centerline{\epsfbox{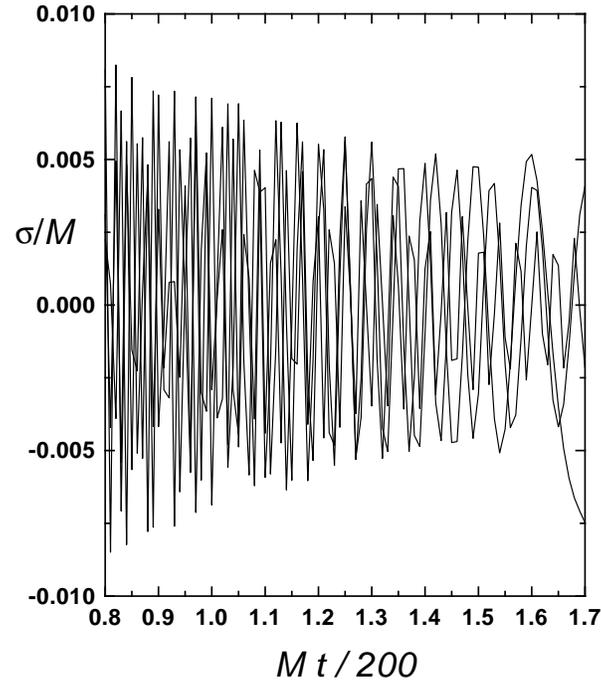}}}
\caption{A blow-out of the initial time evolution of the trajectories
shown in {}Fig. 1.}
%\vspace{1cm} 
 
\end{figure} 

\begin{figure}[b] 
\epsfysize=9cm  
{\centerline{\epsfbox{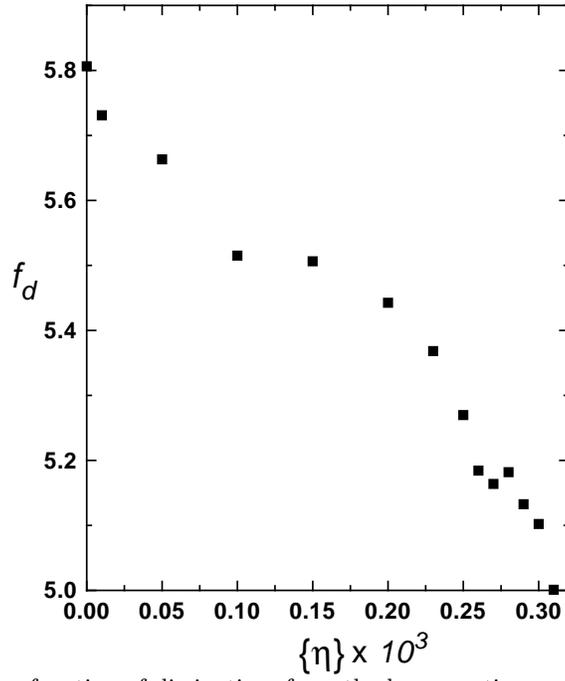}}}
\caption{The fractal dimension as a function of dissipation,
from the box-counting method applied to a box in phase
space centered around the initial conditions used in {}Fig. 1.}
\vspace{1cm} 
 
\end{figure} 

\begin{figure}[b] 
\epsfysize=9cm  
{\centerline{\epsfbox{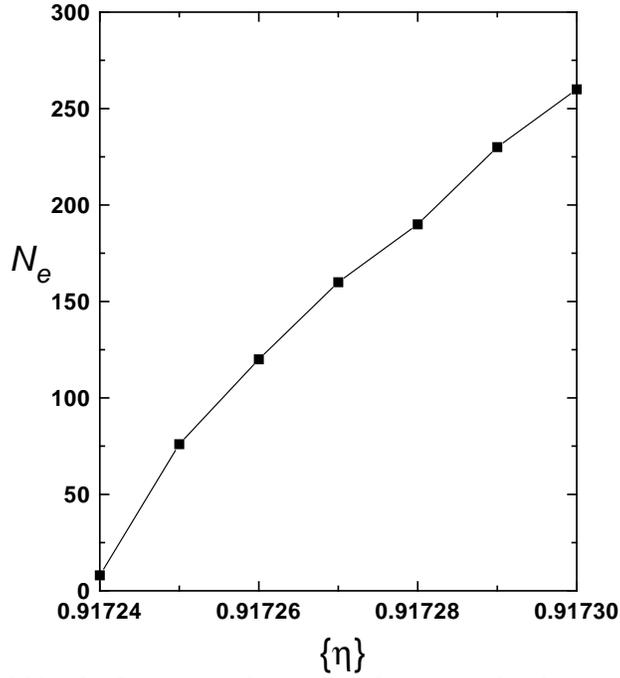}}}
\caption{The total number of e-folds of inflation as a function of 
dissipation for the initial conditions at $t=0$: $\phi=3.5 M$, $\sigma=M$,
$\dot{\phi}= 2.5\times 10^{-2} M^2$, $\dot{\sigma}=5 \times 10^{-4} M^2$
and $\rho_r(t=0)=0$.}
\vspace{1cm} 
 
\end{figure}

\end{document}